# Grid-Connected Photovoltaic System for Active and Reactive Power Management

Ehsan Gholami[1], Arzhang Eradat[1], Hamed Fotoohabadi[2]

[1] Department of Electrical Engineering, Yazd University, Yazd, Iran
[2] Power and Control Department, Islamic Azad University, Shiraz Branch, Shiraz, Iran

*Abstract*—**In this paper, the behavior of a photovoltaic (PV) system, with maximum power point tracking (MPPT) connected to three-phased grid has been investigated. A voltage source inverter (VSI) has been used to connect the photovoltaic system to the grid. The control strategy applied to control inverter switches is based on voltage mode method. Strategy of this project is that during sunlight the system sends active power to the grid and at the same time compensates the reactive power of the load. In case there is no sunlight, the inverter only compensates the reactive of the load. So the photovoltaic system is operated the whole day. In this paper the behavior of the photovoltaic system to provide active power and compensate reactive power of the load is investigated. The time domain simulation is performed using PSCAD/EMTDC software.**

## I. Introduction

The development of the technology in last decades and the need for more electrical power has been led to a global energy problem [1], [2]. Renewable energy sources are a very good solution. Usually, the grid integration of renewable energy sources is carried out by power electronics [3]. The energy generated by photovoltaic systems constitutes a large part of the total amount of energy produced by renewable energy sources. The output power of photovoltaic systems is significantly affected by weather conditions [4]. Therefore extracting maximum power from photovoltaic systems forms a major part of research activities [5]. Several maximum power point tracking algorithms for photovoltaic systems have been developed [6], [7]. These algorithms are applied in DC/DC converters or in DC/AC inverters.

The efficiency of a Photovoltaic system is a serious challenge and many research efforts have been carried out to improve it. These efforts mainly focus on supplying the grid with active and reactive power. In [8] a control algorithm based on synchronous rotating frame (SFR) is proposed. In this method active and reactive power controlled based on the currents in the d-q rotating reference system. In a PV system with a DC-DC converter, two PI controllers and one PLL are used [9]. In case the mathematical model for the control of active and reactive power is not clarified, a DC-AC converter and a PLL are used [10].

A photovoltaic cell can be depicted as an equivalent circuit as shown in fig 1.

The photo current $I_{ph}$ depends on solar radiation $G$ and temperature $T$ of environment according to following equation:

$$I_{ph} = I_{ph}(T_{ref}) \cdot (1 + K_0 \cdot (T - T_{ref}))$$

In this equation, $I_{ph(T_{ref})}$ is the photo current corresponding to the nominal temperature $T_{ref}$.

$$I_{ph}(T_{ref}) = \frac{G}{G_{ref}} \cdot I_{SC}(T_{ref})$$

$$K_0 = \frac{I_{SC}(T) - I_{SC}(T_{ref})}{T - T_{ref}}$$

In these equations, $G_{ref}$ is the nominal radiation given by PVs constructor and $I_{sc}$ is the short circuit current. Diode current, $I_D$ can be presented as:

$$I_D = I_0 \cdot \left[\exp\left(\frac{V_{cell} + I_{cell} \cdot R_S}{V_T}\right) - 1\right]$$

Where $V_{cell}$ and $I_{cell}$ are the output voltage and current for a single PV cell respectively. $V_T$ is the thermal voltage, $I_0$ is diode saturation current and $R_S$ is series resistance. According to ohm's law, current $I_{SH}$ through shunt resistance, $R_{SH}$ is equal to:

$$I_{SH} = \frac{V_{cell} + I_{cell} \cdot R_S}{R_{SH}}$$

By using these equations, $I_{cell}$ can be obtained as follows:

$$I_{cell} = I_{ph} - I_0 \left[\exp\left(\frac{V_{cell} + I_{cell} \cdot R_S}{V_T}\right) - 1\right] - \left(\frac{V_{cell} + I_{cell} \cdot R_S}{R_{SH}}\right)$$

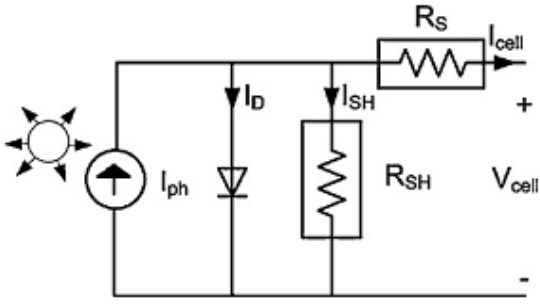

Fig. 1

For $N_s$ in series and $N_v$ in parallel cells, the output voltage $V$ and the output current $I$ can be expressed as:

$$V = N_S \cdot V_{cell}$$
$$I = N_p \cdot I_{cell}$$

## II. MPPT ALGORITHM

In this paper, the incremental conductance (IC) algorithm is used as the MPPT algorithm. Using this, the PV produces maximum power under different conditions of solar radiation and environmental temperature. The IC algorithm is based on the differentiation of PV power and on condition of zero slope of PV curve. The MPP point has been shown at fig. 2.

By comparing the incremental conductance $dI/dV$ to the instantaneous inductance $I/V$, the MPP can be tracked. When MPP has been reached, the operation of PV will be hold at this point. If any changes in $dI$ occur, the algorithm changes $V$ to achieve new MPP point.

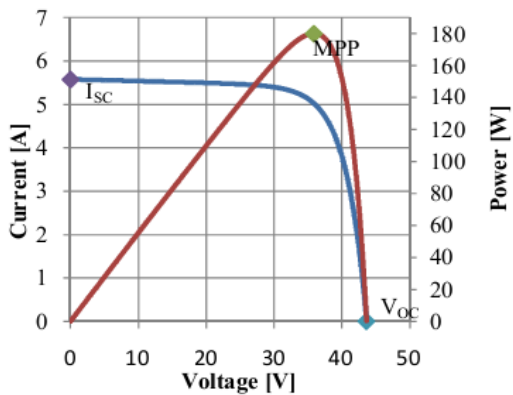

Fig. 2

## III. CONTROL VOLTAGE STRATEGY

In this paper the method is used to control inverter is based on the Voltage Mode Control. The input variables in this strategy are the inverter's input voltage, $V_{in}$ and the reactive power [11], [12]. To generate fire signals of the inverter's switches, first $V_{in}$ reduced from open circuit voltage of the PV, $V_{oc}$ and then it is given to a PI controller. The output will be considered as phase signal. Same process is applied to reactive power to obtain magnitude signal. The process to obtain phase signal and magnitude signal are shown in fig. 3. These two signals are used to generate a sinusoidal signal.

Since the system is used as case study in this paper is a three-phase grid, the output sinusoidal signal must be shifted for each phase. The output signals called carrier signals, finally compared with a high frequency triangle signal to generate proper triggering signals as shown in fig. 4.

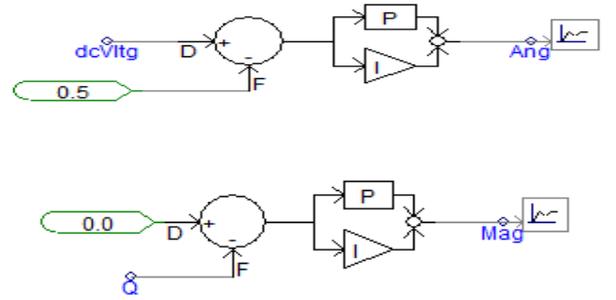

Fig. 3

## IV. CASE STUDY

The system used in this paper to investigate behavior of a PV system during a change in solar radiation is shown in fig. 5. It includes a PV farm connected to a three-phase grid through a three-phase inverter. The parameters of the photovoltaic system are given in table I.

Table I. Parameters of PV farm

| Description | Parameter |
|---|---|
| Number of parallels PV arrays | 2 |
| Number of PV arrays in series | 40 |
| Open circuit voltage | 21.75 V |
| Short circuit current | 3.45 A |
| Reference solar radiation | 1000 W/$m^2$ |
| Reference temperature | 25 °C |
| Saturation diode current | 4.05 e-7 A |

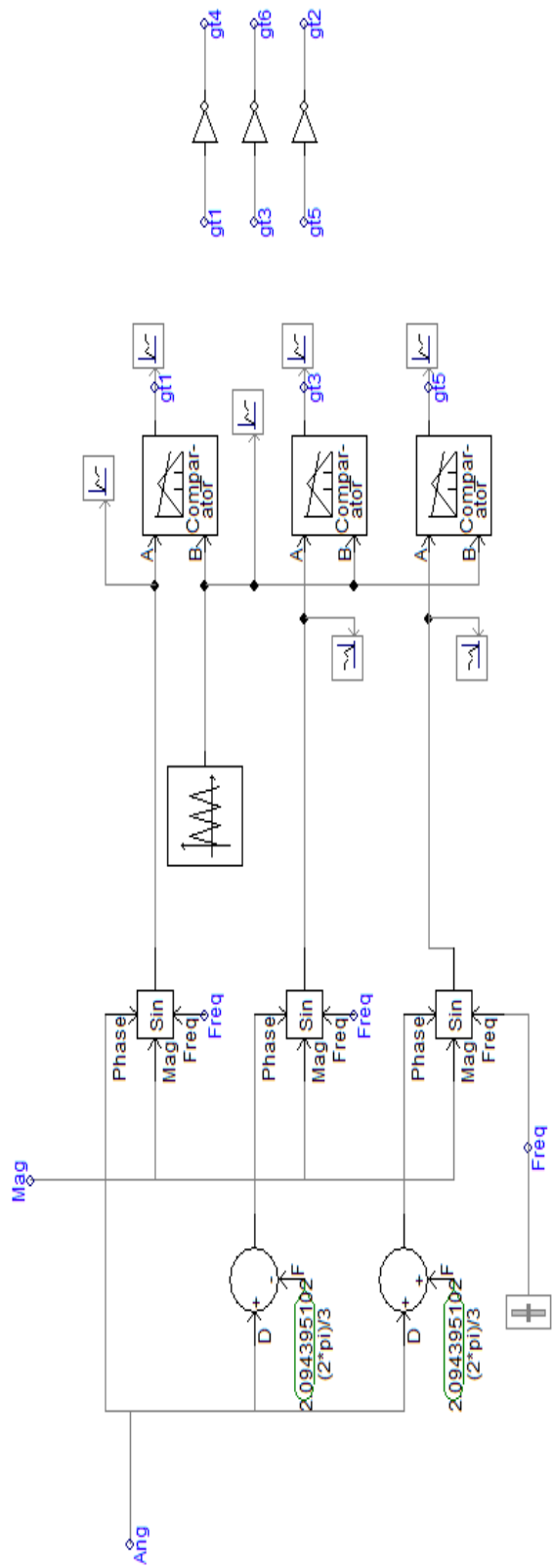

Fig. 4

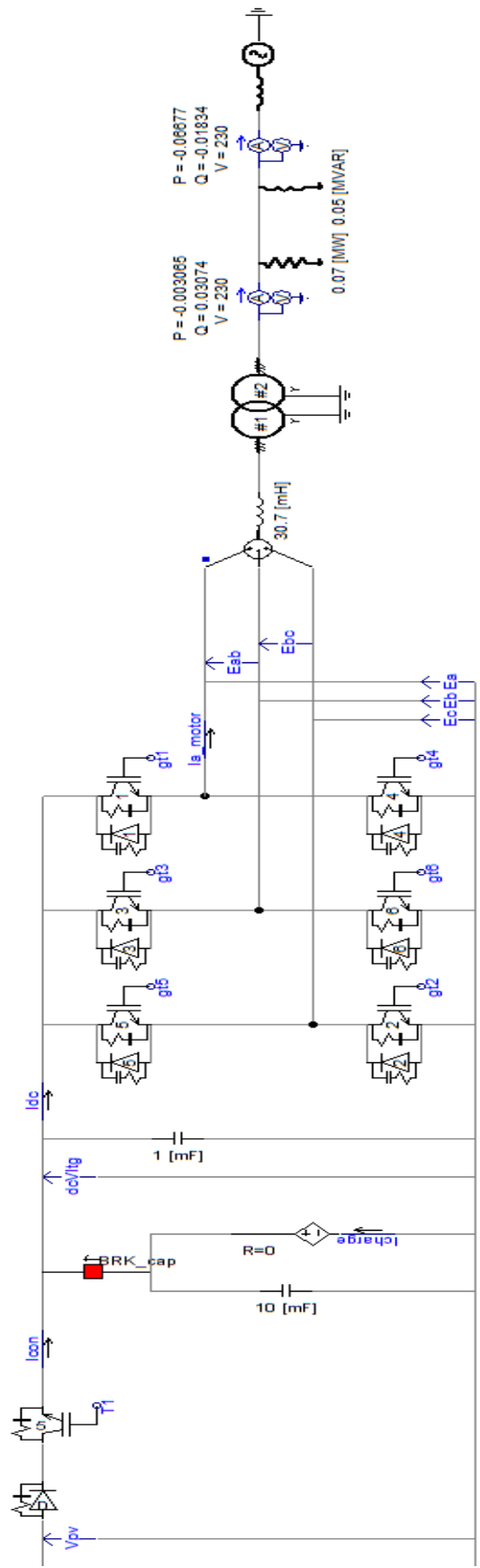

Fig. 5

## V. SIMULATION RESULTS

The time domain simulation has been carried out by PSCAS/EMTDC software. Fig. 6 shows a change in solar radiation at a specific time. To investigate the case of suddenly shading of PV arrays, it is assumed that changes in solar radiation are very fast. Fig. 7 shows corresponding changes in active power of both grid, $P_{network}$ and photovoltaic system, $P_{pv}$. It can be seen from Fig. 7 that during this situation, the generated active power generated by PV system, $P_{pv}$ has been decreased. Therefore the grid active power $P_{network}$ must be increased to supply the load power, $P_{load}$.

Fig .8 shows the reactive power of grid and PV system during the same sudden changes in solar radiation. As it can be seen the PV system acts as a reactive power compensator in case there is no sunlight.

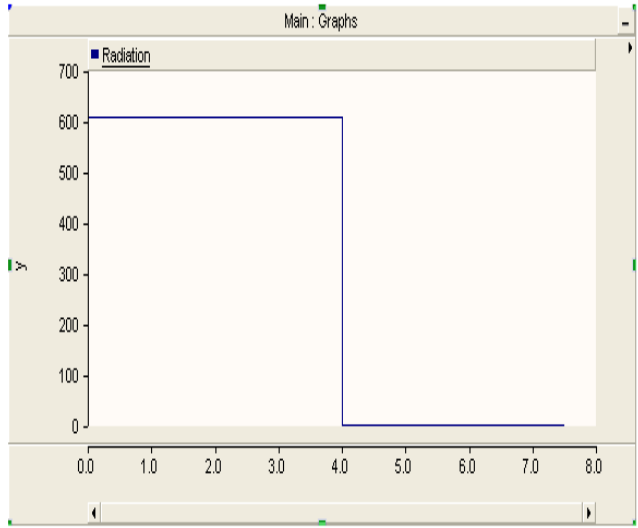

Fig. 6

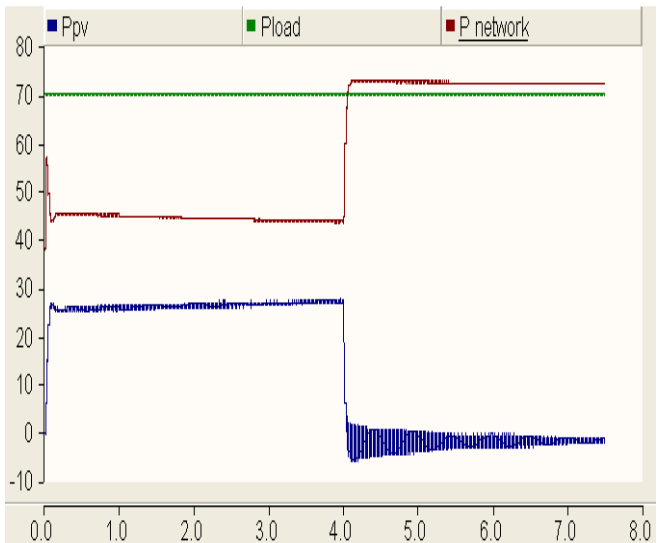

Fig. 7

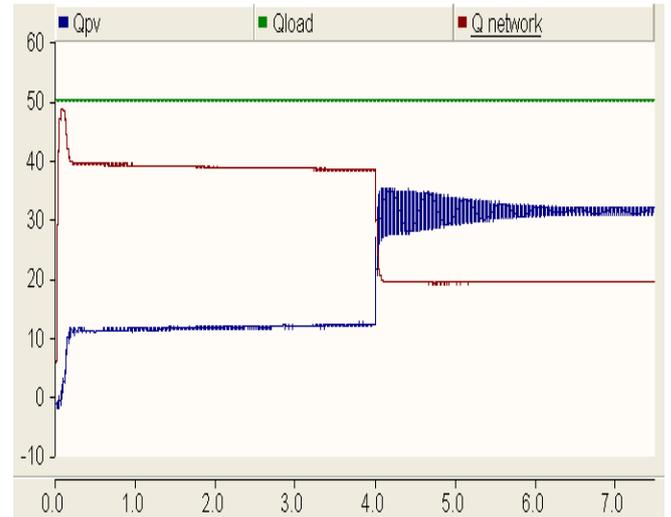

Fig. 8